\documentclass[twocolumn,3p,times]{elsarticle}

\usepackage{amsmath}
\usepackage{amssymb}
\usepackage{bm}
\usepackage{graphicx} 
\usepackage{epstopdf}
\usepackage{enumerate}
\usepackage{booktabs} 
\usepackage{multirow}
\usepackage{threeparttable}
 
\usepackage{color} %\textcolor{}{}
\usepackage{float} 
\usepackage{subfigure}
\usepackage{hhline}

\newtheorem{definition}{Definition}

\newtheorem{theorem}{Theorem}

\newtheorem{remark}{Remark}

\newtheorem{problem}{Problem}
\newtheorem{algorithm}{Algorithm}
\newenvironment{Proof}{{\noindent\it Proof.}\quad}{\hfill $\square$\par}

\usepackage{appendix}

\usepackage{caption}
\biboptions{square, comma, sort&compress, numbers}

\usepackage{paralist}

\begin{document}
\begin{frontmatter}
\title{Space-and-time-synchronized simultaneous vehicle tracking/formation using \\ cascaded prescribed-time control}

\author[NJUST]{Peng Wang }\ead{benjywp711@gmail.com} 
\author[HEU]{Ziyin Chen}\ead{chenziyin\_heu@163.com} 
\author[NJUST]{Xiaobing Zhang \corref{cor} }\ead{zhangxb680504@163.com}         

\cortext[cor]{Corresponding author}
\address[NJUST]{School of Energy and Power Engineering, Nanjing University of Science and Technology, Nanjing 210094, China}                                        
\address[HEU]{Beijing Institute of Space Mechanics and Electricity, China Academy of Space Technology, Beijing 100093,
China}                                        

\begin{abstract}
In this paper, we present a space-and-time-synchronized control method with application to the simultaneous tracking/formation. In the framework of polar coordinates, through correlating and decoupling the reference/actual kinematics between the self vehicle and target, time and space are separated, controlled independently. As such, the specified state can be achieved at the predetermined terminal time, meanwhile, the relative trajectory in space is independent of time. In addition, for the stabilization before the predesigned time, a cascaded prescribed-time control theorem is provided as the preliminary of vehicle tracking control. The obtained results can be directly extended to the simultaneous tracking/formation of multiple vehicles. Finally, numerical examples are provided to verify the effectiveness and superiority of the proposed scheme.
\end{abstract}


\end{frontmatter}

\section{Introduction}
In practice, positioning and tracking problems are of special interest to many important applications which involve the trajectory/path following of autonomous vehicles \cite{aguiar2007trajectory,BWang2021leader},  formation/containment control of multiple autonomous vehicles \cite{WANG201826,liu2018event} and so on. 
It is noted that there exist two important scales in positioning and tracking: space and time. Existing researches usually focus on when to reach a designated location, or what kind of trajectory in space, but few considerations involve both of them at the same time, which inspires this work. 

First of all, a concept named 'synchronization of space and time' is proposed.

\emph{Synchronization of space and time:}
 We summarize the following three characteristics for the synchronization of space and time:
\begin{enumerate}[(i)]
\item  Time and space are separated and can be controlled independently;
\item  In the time control, the specified position can be reached at the predetermined instant;
\item  In the space control, the trajectory in space is independent of time, that is, the trajectory in space is fixed regardless of the terminal time.
\end{enumerate}

In the following, two aspects are discussed.

\emph{Prescribed-time/distance control:} Finite/fixed /prescribed-time control has drawn  increasing attention in recent years in view of the fact that many practical applications require severe time response constraints.  Homogeneity property gives a useful analysis for HOSM (high-order sliding mode) control, and if an asymptotically stable system is homogeneous of a negative degree, then it is finite-time stable to a sliding manifold (see \cite{Davila13,Levant05c,Bhat05}).
However, the settling time in these finite-time methods grows unbounded when initial conditions tend to infinity, and the initial conditions are hard
to acquire in some practical situations \cite{Hu2019concurrent}.
Fixed-time stability requires a controller (observer) to provide some desired control (observation) precision at a given time, independent of the initial conditions, and  the settling time is subject to an upper bound but varies due to uncertainties and nonlinearities
 \cite{Tian17,Polyakov15}. 
 The reviews upon the fixed-time stability and its applications can be found in \cite{Zuo2018overview}.  
 However, the fixed-time settling time can not be preassigned arbitrarily since its upper bound is subject to certain restrictions \cite{YWang2018}. 
Moreover, prescribed-time control employs a scaling of state by a function of time that grows unbounded to the terminal time and yields the regulation in prescribed finite time(see \cite{song2017time,wang2018leader}).
Furthermore, regarding the prescribed-distance control problem, the system regulation can be achieved before the prescribed distance is reached.  
The conversion between the prescribed-distance control and the existing prescribed-time control only requires the distance-driven transformation for the differential dynamic system model, which will be answered in this paper.

\emph{Simultaneity in time: }The finite-time distributed consensus has been a popular topic in recent years, which is known as one of the most important issues in cooperative control\cite{cao2014finite,zhao2017adaptive,wang2018leader}. However, for some classes of practical systems, this finite- time convergence is not enough operating, for example, for a service robot hand, its joints are usually required to reach the desired angle at the same time; the proximity operations require that the pursuer perform the translational and rotational maneuvers with respect to the target simultaneously \cite{shao2020adaptive}; for the multi-missile interception active defense \cite{zhou2016distributed,CHEN2019454}, multiple missiles are expected to strike the target simultaneously. In this regard, simultaneity in time for multi-agent systems are of great importance.
The works in \cite{DLi21Ontime,DLi21Fixedtime} proposed the
time-synchronized
consensus control, focusing on how to design a controller which drives
all the elements of all the agent states to consensus at the
same time; and fixed-time-synchronized consensus, where the
upper bound of the synchronized settling time is independent of the initial states of multi-agent systems.

\emph{Contributions of this work:} The main contributions are stated as following
\begin{enumerate}[(i)]
\item  {In the framework of polar coordinates, given a reference relative trajectory, the developed distance-driven controller enables the prescribed-distance stability and achieves the separation between space and time. In comparison, few of the existing works consider the distance-driven transformation about the tracking problem. Moreover, the work in \cite{CHEN2019454} employs a different distance-driven transformation and obtains the travelling range in advance, while  the velocity value is set fixed and the target is assumed stationary;}
\item Through correlating and decoupling the reference/actual kinematics between the self vehicle and target, the control variables related to velocity and inclination angle are controlled respectively to form synchronization between the arrival time and path trajectory. {This kind of correlating and decoupling operations give the freedom to the velocity control in polar coordinates, which is proposed for the first time, to the best of the authors’ knowledge};
\item  A cascaded prescribed-time control theorem is proposed which enables the prescribed-time stabilization for the general strict-feedback systems. Compared with the prescribed-time control \cite{song2017time,wang2018leader},  the proposed theorem eliminates restrictions on the system model and allows the uncertainties and the remaining-time-related interconnections in each subsystem; the cascaded prescribed-time tracking problem can be directly solved through the separately prescribed-time control design of each subsystem;
\item The synchronization of space and time control enables the simultaneous arrival of multiple vehicles which reach the designated position or achieve formation simultaneously.
{Compared with the existing simultaneous control work, the target in the simultaneous tracking/formation problem does not have to be set stationary as in \cite{zhou2016distributed,CHEN2019454}, and the simultaneous property can be directly derived from the proposed space-and-time-synchronized control.}
\end{enumerate}

\emph{Notation}:
Throughout this paper, ${\mathcal{R}}$ denotes the set of real numbers, ${\mathcal{R}}^{n\times m}$ represents the set of $n \times m$ real matrix, $\mathcal{N}_+$ is the set of positive integers.
% $I_n$ denotes a $n$-dimensional identity matrix of appropriate dimension, $\bold{1}_n$ is a $n$-dimensional vector with all elements equal to 1. %Additionally, $I$ represents the identity matrix with compatible dimensions.
 %Additionally, $I$ and $O$ represent the identity matrix and zero matrix with compatible dimensions, respectively. 
$( \cdot )^T$ is the transpose of one matrix, 
%sgn$( \cdot )$ denotes the standard sign function, 
diag$( \cdot )$ represents the diagonal matrix. $\left|  \cdot  \right|$ is the absolute value of a scalar and $\left\|  \cdot  \right\|$ is defined as the Euclidean norm of a vector. Given a matrix $M$, $M>0$ (or $M<0$) means that $M$ is a positive definite (or negative definite) matrix, $\lambda_\text{min}(M)$ and $\lambda_\text{max}(M)$ denote the the minimum and maximum eigenvalue of matrix $M$, respectively.

\section{Formulation of space-and-time-synchronized problem}
\begin{figure}[!t]
\centerline{\includegraphics[width=0.4\textwidth]{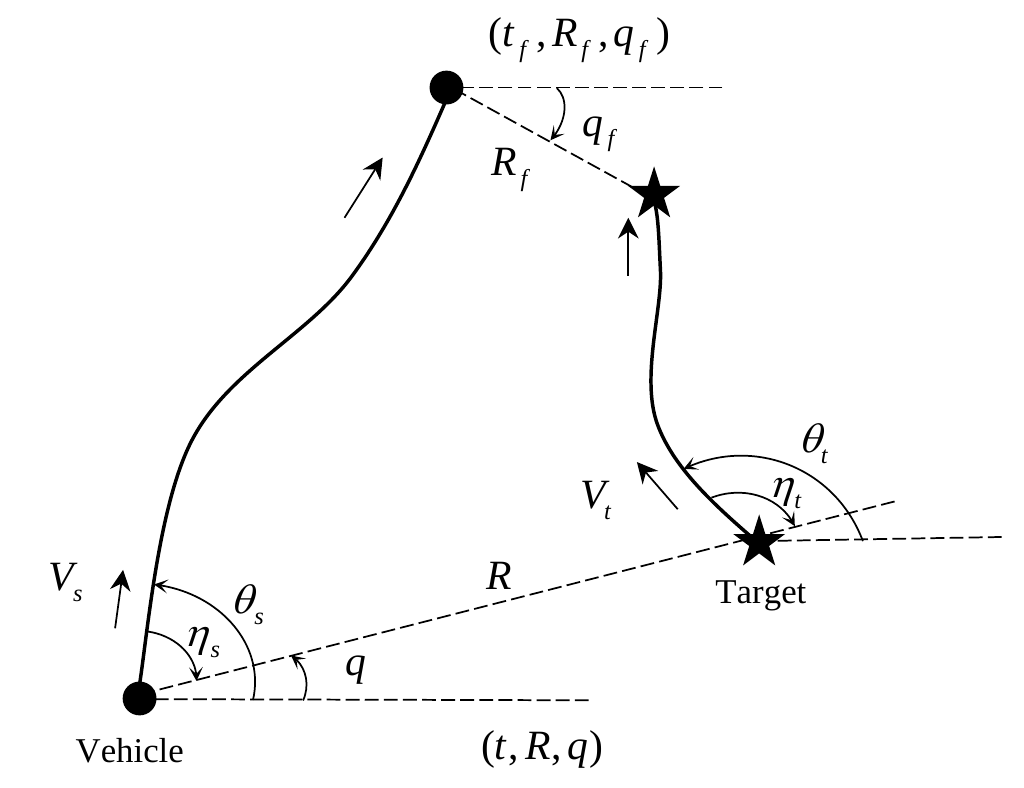}}
\caption{Geometry description}
\label{fig:Fig_geometry}
\end{figure}

As the geometry of planer scenario depicted in Fig. \ref{fig:Fig_geometry}, the subscripts $s$ and $t$ denote the self vehicle and target, respectively, the subscripts $0$ and $f$ represent the initial time and the predetermined time, respectively.  $R \in \mathcal{R}$ represents their relative distance along the line of sight (LOS) angle $q \in \mathcal{R}$. ${{V}_{s}}\in \mathcal{R}$ and ${{V}_{t}} \in \mathcal{R}$ denote the velocities of the vehicle and target, respectively, ${{\theta }_{s}} \in \mathcal{R}$ and ${{\theta }_{t}} \in \mathcal{R}$ are their corresponding path angles, ${{\eta }_{s}} \in \mathcal{R}$ and ${{\eta }_{t}} \in \mathcal{R}$ represent the heading angles. The tracking kinematics between the vehicle and target in the framework of polar coordinates are modeled as
\begin{equation}\label{Eq_2_1}
\begin{aligned}
  & \dot{R}=-{{V}_{s}}\cos {{\eta }_{s}}+{{V}_{t}}\cos {{\eta }_{t}}, \\ 
 & R\dot{q}={{V}_{s}}\sin {{\eta }_{s}}-{{V}_{t}}\sin {{\eta }_{t}}, \\ 
 & {{\eta }_{s}}=q-{{\theta }_{s}},\ {{\eta }_{t}}=q-{{\theta }_{t}}  
\end{aligned}
\end{equation}
with the initial condition ${{R}}(0)=R_0$,  ${{q}}(0)=q_0$, ${{\eta}}_s(0)=\eta_{s0}$ and  ${{V}}_s(0)=V_{s0}$. When it comes to a special case $V_t=0$ which means a stationary target point, the kinematics model (\ref{Eq_2_1})  is reduced to a positioning problem.

Define the pair $(t,R, q)$: at time $t$, the distance and LOS angle between the vehicle and target are $R$ and $q$, respectively. We can now formulate the problem as follows: 

\begin{problem} \label{Problem_2_1}
(Synchronization of space and time) For a vehicle tracking problem (\ref{Eq_2_1}), how to ensure the vehicle starting from $(t_0,R_0, q_0)$ eventually realizes $(t_f,R_f,q_f)$ and gets maintained in $(t,R_f,q_f)$ for $t>t_f$?
\end{problem}
 
\begin{algorithm}
For the sake of generality, the solution with respect to Problem \ref{Problem_2_1} can be summarized as 
\begin{enumerate}[(i)]
\item   In the framework of polar coordinates, introduce the reference trajectory $(R_d,q_d)$ for the vehicle-target relative motion, which achieves the separation of time and space, see Section II.A;
\item   Correlate and decouple the reference/actual relative kinematics  to obtain the separated control variables related to velocity $V_s$ and inclination angle $\eta_s$, see Section II.B;
\item   The travelling  time and path trajectory are controlled respectively to form synchronization between space and time.
\end{enumerate}
\end{algorithm}
\subsection{Reference relative tracking trajectory }
The motion equations for the reference tracking problem are given by
\begin{equation}\label{Eq_2_2}
\begin{aligned}
  & {{{\dot{R}}}_{d}}=-{{V}_{d}}\cos {{\eta }_{d}}, \\ 
 & {{R}_{d}}{{{\dot{q}}}_{d}}={{V}_{d}}\sin {{\eta }_{d}} \\ 
\end{aligned}
\end{equation}
where $R_d$ and $q_d$ denote the distance and LOS angle of the reference trajectory, respectively, $V_d$ and $\eta_d$ are the corresponding velocity and heading angle, respectively. Their initial conditions are set ${{R}_{d}}(0)=R_0$,  ${{q}_{d}}(0)=q_0$, ${{\eta}_{d}}(0)=\eta_{d0}$ and ${{V}_{d}}(0)=V_{d0}$.

Define a new distance-scaled variable ${{{r}}_{d}}$ and a prescribed value ${{{r}}_{f}}$ 
\begin{align} \label{Eq_2_3}
&{{{r}}_{d}}=
\ln (R_0-{R}_{f1})-\ln ({{R}_{d}}(t)-{R}_{f1}), \ \ R_d \in \left( R_f,R_0  \right]    ,  \notag \\
&{{{r}}_{f}}=\ln (R_0-{R}_{f1})   
\end{align}
where ${R}_{f1}=R_f-1$. It is obvious that $r_d=r_f$ when $R_d=R_f$. 
The relationship between the distance-scaled variable ${r}_d$ and the actual time $t$ is subject to

\begin{equation}\label{Eq_2_4}
{{{{\dot{r}}}_{d}}}=
{{{V}_{d}}\cos {{\eta }_{d}}}/({{{R}_{d}}-{R}_{f1}}), \ \ R_d \in \left( R_f,R_0  \right].    
\end{equation}
%\begin{remark} \label{Remark_2_1} 
%It is obvious that $r_f-r_d=0$ when $R_d=R_f+1$. For a practical tracking scenario with $R_f \gg 1$ or $V_{df} \gg 1$, $R_d \in \left(R_f,R_f+1 \right)$ is acceptable. Otherwise, the distance-scaled variable ${r}_d$ can be redefined by ${{{r}}_{d}}=\ln (R_0-R_f)-\ln ({{R}_{d}}(t)-R_f+1)$.
%\end{remark}
 
Taking derivative with respect to the distance-scaled variable $r_d$, rather than the time variable $t$, yields the rewritten dynamics: {when $r_d \in \left[0, r_f \right)$,}
\begin{align} \label{Eq_2_5}
  & \frac{\text{d}{{q}_{d}}}{\text{d}{{{r}}_{d}}}=\frac{{{R}_{d}}-{R}_{f1}}{{{R}_{d}}}\tan {{\eta }_{d}}, \\ 
 & \frac{{{\text{d}}^{2}}{{q}_{d}}}{\text{d}r_{d}^{2}}=\frac{R_d-{R}_{f1}}{{{V}_{d}}{{\cos }}{{\eta }_{d}}}\left[\frac{R_d-{R}_{f1}}{R_d \cos^2{\eta_d}}\frac{\text{d}{{\eta }_{d}}}{\text{d}t}
- \frac{{R}_{f1} V_d \sin{\eta_d}}{R_d^2}
 \right]. \notag
\end{align}

\begin{theorem} \label{Theorem_1}
Consider the system in form of (\ref{Eq_2_2}).

(i)  When $R_d \in  [0, r_f  )$,  the control law of $\eta_d$ is designed according to (\ref{Eq_2_6}). The system state pair $(q_d,\eta_d)$  asymptotically converges to the desired value $(q_f,0)$ before ${{r}_{d}}$ approaching the prescribed value ${{r}_{f}}$.
\begin{align} \label{Eq_2_6}
{\dot{\eta }_{d}}=&-\frac{R_d{{V}_{d}}{{\cos }^{3}}{{\eta }_{d}}}{({{R}_{d}}-{R}_{f1})^2}\left[ \frac{k_{d2}}{{{{r}}_{f}}-{{{r}}_{d}}}\frac{\text{d}{{q}_{d}}}{\text{d}{{{r}}_{d}}}+\frac{k_{d1}({{q}_{d}}-{{q}_{f}})}{{{({{{r}}_{f}}-{{{r}}_{d}})}^{2}}}  \right]  \notag \\
& + \frac{{R}_{f1} V_d \cos^2{\eta_d}\sin{\eta_d}}{(R_d-{R}_{f1})R_d}. 
\end{align}
When letting 
${{k}_{d1}}=N_1N_2, \  {{k}_{d2}}=N_1+N_2-1$, $N_1<N_2$ and
$N_1, N_2 \in \mathcal{N}_+$, the travel range can be calculated as
\begin{equation} 
\begin{aligned}
S_d=\int_{R_f}^{{{R}_{0}}}{\sqrt{1+{{(\tan {{\eta }_{d}})}^{2}}}}\text{d}R_d .
\end{aligned}
\end{equation}
In this sequel, $S_d$ is irrespective of $V_{d}$, the travel time $t_{f}$ is only related to the velocity $V_{d}$. 
For any constant velocity ${{V}_{d}}$, the travel time is ${{t}_{f}} ={S_d}/{V_d}$.

(ii) When $R_d$ approaches $R_f$, through $\dot{V}_{d}=0$ and  $\dot{\eta}_d=0$, the system state pair $(R_d,q_d,\eta_d)$  gets maintained in the desired value $(R_f,q_f,0)$.
\end{theorem}

\begin{Proof}
See Appendix A.
\end{Proof}

\begin{remark} 
Here comes the motivation of introducing the proposed distance-scaled variable $r$ and doing  the $r$-related dynamics transformation:
\begin{enumerate} [(i)]
\item The system dynamics (\ref{Eq_2_5}) with respect to the introduced distance-scaled variable $r_d$ is independent of time and directly linked to the terminal distance $R_f$. As a result, this proposed distance-related transformation is more intuitive to the distance tracking problem like the prescribed-distance control;
\item This kind of transformation (\ref{Eq_2_3}) together with the controller (\ref{Eq_2_6}) enables the achievement of regulation objective before $r_d=r_f$. The resultant irrelevance between $S_d$ and $V_{d}$ creates a separation between space and time, which provides the possibility for the space-and-time-synchronized control;
\item Such a state transformation brings $\eta_d$ into a state-related form, i.e. $\tan \eta_d=\frac{\text{d}{{q}_{d}}}{\text{d}{{{r}}_{d}}} \frac{R_d}{R_d-R_{f1}}$, such that the specific expression $\eta_d$ in space can be directly obtained in (\ref{Eq_2_11}). Some heading-angle-constrained problems can be handled through appropriate parameter selection when considering the field-of-view limits.   
\end{enumerate}
\end{remark}

\begin{figure*}[htbp]
%\hrulefill
\normalsize
\begin{equation*}  
\begin{aligned}
& M=\left[ \begin{matrix}
   -\cos {{\eta }_{d}} & {{V}_{s}}\left( \cos {{\eta }_{d}}\int_{0}^{1}{\sin (\tau {{\eta }_{e}})\text{d}\tau }+\sin {{\eta }_{d}}\int_{0}^{1}{\cos (\tau {{\eta }_{e}})\text{d}\tau } \right)  \\
   \sin {{\eta }_{d}} & {{V}_{s}}\left( \cos {{\eta }_{d}}\int_{0}^{1}{\cos (\tau {{\eta }_{e}})\text{d}\tau }-\sin {{\eta }_{d}}\int_{0}^{1}{\sin (\tau {{\eta }_{e}})\text{d}\tau } \right)  \\
\end{matrix} \right], \ \ G=\left[ \begin{matrix}
   {{V}_{t}}\cos {{\eta }_{t}}  \\
   -{{V}_{t}}\sin {{\eta }_{t}}-{{V}_{d}}\sin {{\eta }_{d}} ({{{R}_{e}}}/{R_d} ) \\
\end{matrix} \right].
\end{aligned}
\end{equation*}
\hrulefill 
%\vspace*{4pt} 
\end{figure*}

\subsection{Kinematic modelling}
To correlate the reference/actual relative kinematics, we define the following error variables, i.e., distance error ${{R}_{e}}=R-{{R}_{d}}$, LOS angle error ${{q}_{e}}=q-{{q}_{d}}$, angle ${{\eta }_{e}}={{\eta }_{s}}-{{\eta }_{d}}$, velocity error ${{V}_{e}}={{V}_{s}}-{{V}_{d}}$.

Taking time derivative of ${{R}_{e}}$ along the solutions of (\ref{Eq_2_1}) and (\ref{Eq_2_2}) leads to

\begin{align}\label{Eq_2_16}
 {{{\dot{R}}}_{e}} & =-{{V}_{s}}\cos {{\eta }_{s}}+{{V}_{s}}\cos {{\eta }_{d}}-{{V}_{s}}\cos {{\eta }_{d}}\nonumber \\
 & \ \ \ +{{V}_{d}}\cos {{\eta }_{d}}+{{V}_{t}}\cos {{\eta }_{t}} \nonumber\\ 
 & =-{{V}_{s}}(\cos ({{\eta }_{d}}+{{\eta }_{e}})-\cos {{\eta }_{d}})-{{V}_{e}}\cos {{\eta }_{d}}+{{V}_{t}}\cos {{\eta }_{t}} \nonumber\\ 
 & =-{{V}_{e}}\cos {{\eta }_{d}}+{{V}_{t}}\cos {{\eta }_{t}} \nonumber \\
 & \ \ \ +{{V}_{s}}\left[ \cos {{\eta }_{d}}\frac{1-\cos {{\eta }_{e}}}{{{\eta }_{e}}}+\sin {{\eta }_{d}}\frac{\sin {{\eta }_{e}}}{{{\eta }_{e}}} \right]{{\eta }_{e}}. 
\end{align}

A similar operation for ${{q}_{e}}$ brings 
\begin{align}\label{Eq_2_17}
  R{{{\dot{q}}}_{e}} =&{{{V}_{s}}\sin {{\eta }_{s}}}-{{{V}_{s}}\sin {{\eta }_{d}}}+{{{V}_{s}}\sin {{\eta }_{d}}}  -{{{V}_{d}}\sin {{\eta }_{d}}} \nonumber\\
  &  -{{{V}_{d}}\sin {{\eta }_{d}}}({R_e}/{R_d})-{{{V}_{t}}\sin {{\eta }_{t}}} \nonumber\\ 
   =&{{{V}_{e}}\sin {{\eta }_{d}}}+
{V_s} \left[ \sin {{\eta }_{d}}\frac{\cos {{\eta }_{e}}-1}{\eta_e}+\cos {{\eta }_{d}} \frac{\sin {{\eta }_{e}}}{\eta_e} \right]{\eta_e}   
 \nonumber\\ 
 & -{{{V}_{d}}\sin {{\eta }_{d}}}({R_e}/{R_d})-{{{V}_{t}}\sin {{\eta }_{t}}} .
\end{align}

Invoking the fact
\begin{equation} \label{Eq_2_19a}
\begin{aligned} 
& \frac{1-\cos {{\eta }_{e}}}{{\eta }_{e}}=\int_{0}^{1}{\sin (\tau {{\eta }_{e}})\text{d}\tau },  \ \ \frac{\sin {{\eta }_{e}}}{{\eta }_{e}}=\int_{0}^{1}{\cos (\tau {{\eta }_{e}})\text{d}\tau }   
\end{aligned} 
\end{equation}
and combing the above dynamics of ${{R}_{e}}$ and ${{q}_{e}}$, we readily obtain the kinematic model of vehicle tracking problem
\begin{equation}\label{Eq_2_19}
\begin{aligned}
\left[ \begin{matrix}
   {{{\dot{R}}}_{e}}  \\
   R{{{\dot{q}}}_{e}}  \\
\end{matrix} \right]= &M({{\eta }_{s}},{{\eta }_{e}},V_s)\left[ \begin{matrix}
   {{V}_{e}}  \\
   {{\eta }_{e}}  \\
\end{matrix} \right] +G(R_e,R_d,{{V}_{t}},{{\eta }_{t}},{{V}_{d}},{{\eta }_{d}})
\end{aligned}
\end{equation}
where the detailed expressions of $M({{\eta }_{s}},{{\eta }_{e}},V_s)$ and $G(R_e,R_d,{{V}_{t}},{{\eta }_{t}},{{V}_{d}},{{\eta }_{d}})$ are at the top of next page.

\begin{remark}
With regard to (\ref{Eq_2_19a}), a special case is when $\eta_e=0$, { through the L'Hospital rule and limit of indeterminate form, one obtains $(1-\cos \eta_e)/\eta_e=0$ and $\sin \eta_e /\eta_e=1$.}
\end{remark}

\begin{remark}
As for the parameter matrix $M({{\eta }_{s}},{{\eta }_{e}},V_s)$, its determinant is {
$\left| M({{\eta }_{s}},{{\eta }_{e}},V_s) \right|= -{{V}_{s}} (\sin \eta_e /\eta_e)<0
$},
which contributes to the full rank of $M({{\eta }_{s}},{{\eta }_{e}},V_s)$ and enables the later matrix inversion operation.
\end{remark}

\subsection{Dynamic modelling}
Define the control variable $u={{\left[ \begin{matrix}
   {{u}_{V}} & {{u}_{\theta }}  \\
\end{matrix} \right]}^{T}}$  where ${{u}_{V}}$ is the tangential acceleration and ${{u}_{\theta }}$ is the lateral acceleration.
 The vehicle's dynamic model is expressed by
\begin{equation}\label{Eq_2_21}
\begin{aligned}
  & {{{\dot{V}}}_{s}}={{u}_{V}}+d_V ,\\ 
 & {{{\dot{\theta }}}_{s}}={u_\theta}/{{{V}_{s}}} +d_\theta  
\end{aligned}
\end{equation}
where $d_V$ and $d_\theta$ are the external disturbances. 

As such, the compact system model combing (\ref{Eq_2_19}) and (\ref{Eq_2_21}) is written as
\begin{subequations}
\begin{align}
&   \left[ \begin{matrix}
   {\dot{R}_{e}}  \\
   {\dot{q}_{e}}  \\
\end{matrix} \right]=  \bar{M} \left[ \begin{matrix}
   V_s \\
    {\eta}_s   \\
\end{matrix} \right]- \bar{M} \left[ \begin{matrix}
   V_d \\
   \eta_d    \\
\end{matrix} \right]+\bar{G},   \label{Eq_3_2} \\
&
 \left[ \begin{matrix}
   {{{\dot{V}}}_{s}}  \\
   {{{\dot{{\eta }}}}_{s}}  \\
\end{matrix} \right]= {B}\left[ \begin{matrix}
   {{u}_{V}}  \\
   {{u}_{\theta }}  \\
\end{matrix} \right]+ H+ {F}  \label{Eq_3_3}
\end{align}
\end{subequations}
where 
$ \bar{M}(R,{{\eta }_{s}},{{\eta }_{e}},V_s)=\text{diag}(1,1/R)M$,  
$ \bar{G}(R,R_e,R_d,{{V}_{t}},{{\eta }_{t}},{{V}_{d}},{{\eta }_{d}})=\text{diag}(1,1/R)G$ ,  
$ B(V_s)=\text{diag}(1,-1/V_s)$ ,  
$  H(\dot{q},\dot{R},R,\eta_s)=\left[ \begin{matrix}
   0   &
   \dot{q}   
\end{matrix} \right]^T,  {F}(d_V,d_\theta)= \left[ \begin{matrix}
   d_V   &
   -d_\theta   
\end{matrix} \right]^T$ .

\section{Prescribed-time Control Design}
For the prescribed-time control design, we reach the following consensus:
as long as $R_e$ and $q_e$ converge to 0 before the final time $t_f$ (before $R$ converges to ${R}_{f}$), the vehicle $(R,q)$ reaches the specified state $(R_f,q_f)$ at the time $t_f$.

\begin{problem} \label{Problem_2}
(Prescribed-time stabilization) For some kind of strict feedback systems like system (\ref{Eq_3_2})(\ref{Eq_3_3}), how to ensure the system stability within the prescribed time $t \in \left[ 0, t_f \right)$ ?
\end{problem}
\begin{algorithm}
The corresponding solution with respect to Problem \ref{Problem_2} is summarized as
 \begin{enumerate}[(i)]
\item   The prescribed-time stability of a single first-order system can be achieved by employing a scaling of the state by a function of $t_f-t$;
\item   During the backstepping design procedure, the above first-order prescribed-time control design is repeated for each subsystem. In this sequel, a cascaded system consisting of two prescribed-time stable subsystems is obtained;
\item   The prescribed-time stability of the cascaded system is addressed and extended to cascade cases of multiple subsystems.
\end{enumerate}
\end{algorithm}
\subsection{Prescribed-time stabilization of cascaded systems}
Before processing the prescribed-time control, a cascaded prescribed-time theorem is presented as a preliminary. 
Cascades-based control essentially uses the designed control law to make the closed-loop system in a cascaded structure, which usually has the advantage of reducing the complexity of the controller and the difficulty of stability analysis.  

Consider a kind of cascaded system described by
\begin{equation}  \label{EQ_Cascaded_N}
\begin{aligned} 
 & {{\Sigma }_{i}}:\dot{x}_i(t)={{f}_{i}}(t,x_i,{{\Delta }_{x,i}})+g_i(t,x_i,x_{i+1})x_{i+1}, \\
  & {{\Sigma }_{n}}:\dot{x}_n(t)={{f}_{n}}(t,x_n,{{\Delta }_{x,n}}),  i=1,\cdots,n-1   
\end{aligned}
\end{equation}
%\begin{equation}\label{Eq_3_4}
%\begin{aligned}
%  & {{\Sigma }_{1}}:\dot{x}_1(t)={{f}_{1}}(t,x_1,{{\Delta }_{x,1}})+g(t,x_1,x_2)x_2, \\ 
% & {{\Sigma }_{2}}:\dot{x}_2(t)={{f}_{2}}(t,x_2,{{\Delta }_{x,2}}) \\ 
%\end{aligned}
%\end{equation}
where $x_i(t) \in {\mathcal{R}^{n_i }}$ is the system state of subsystem ${{\Sigma }_{i}}$. ${{f}_{i}}(t,x_i,{{\Delta }_{x,i}})$, $g(t,x_i,x_{i+1})$ are continuous in their arguments, and locally Lipschitz in $x_i$, $(x_i,x_{i+1})$ respectively. For a simplified presentation, we view the cascaded system as the subsystem ${{\hat{\Sigma }}_{i}}$ perturbed by the output of subsystem ${{\Sigma }_{i+1}}$.
\begin{equation}\label{Eq_3_5}
{{\hat{\Sigma }}_{i}}:\dot{x}_i(t)={{f}_{i}}(t,x_1,{{\Delta }_{x,i}}).
\end{equation}

\begin{definition}
\cite{Panteley01} A continuous function $\alpha :\left[ 0,a \right)\to \left[ 0,\infty  \right)$ belongs to class $\mathcal{K}$ if it is increasing and $\alpha (0)=0$, {if moreover $\alpha (a)\to \infty$ as $a\to \infty $, it belongs to class $\mathcal{K}_\infty$.} A continuous function $\beta :\left[ 0,a \right)\to \left[ 0,\infty  \right)$ belongs to class $\mathcal{L}$ if it is decreasing and $\beta (s)\to 0$ as $s\to \infty $.
Again, a continuous function  $\gamma :\left[ 0,a \right)\times \left[ 0,a \right)\to \left[ 0,\infty  \right)$ belongs to class $\mathcal{KL}$ if the mapping $\gamma (r,s)$ belongs to class $\mathcal{K}$ with respect to $r$ when fixed $s$, and the mapping $\gamma (r,s)$ decreases along $s$, i.e., $\gamma (r,s)\to 0$ as $s\to \infty $, when fixed $r$.
\end{definition}

\begin{definition} \label{definition_2}
Consider the system $\dot{x}=f(t,x)$, where $f(t,x)$ is piecewise continuous in $t$ and locally Lipschitz in $x$.  The system is globally uniformly prescribed-time stable (GUPTS), if there exist a class $\mathcal{KL}$ function $\gamma (\cdot ,\cdot )$ and a settling-time $t_f$ such that for any initial values $x({{t}_{0}})$:
\begin{equation}
\left\| x(t) \right\|\le \gamma (\left\| x({{t}_{0}}) \right\|,\tau(t,t_f) ),\ \ \ {{t}_{0}}\le t< {{t}_{f}} 
\end{equation}
where $\tau(t,t_f) \to \infty$ as $t \to t_f$, from which one obtains the strictly prescribed-time convergence of $x$: $x(t)\to 0$ as $t\to {{t}_{f}}$. 
\end{definition}

\begin{theorem} \label{Theorem_Cascade_N}
Consider the cascaded system   (\ref{EQ_Cascaded_N}) with $n$ subsystems.
The globally uniformly prescribed-time stability (GUPTS) of the subsystem $\Sigma_i$, $i=1,\cdots,n$ can be obtained, if the following assumptions hold:

(i) Assumption on subsystem  ${{\hat{\Sigma }}_{i}}$, $i=1,\cdots, n$: the subsystem ${{\hat{\Sigma }}_{i}}$ is GUPTS, and there exists a continuously differentiable Lyapunov function ${{V}_{i}}(t,x_i)$ such that
\begin{align} \label{assumption_cascaded}
  & {{\alpha }_{i1}}(\tau){{\left\| x_i \right\|}^{2}}\le {{V}_{i}}(t,x_i)\le {{\alpha }_{i2}}(\tau){{\left\| x_i \right\|}^{2}},  \notag \\ 
 & {{{\dot{V}}}_{i}}={\partial {{V}_{i}}}/{\partial t}\;+({\partial {{V}_{i}}}/{\partial x_i}\;){{f}_{i}}(t,x_i,{{\Delta }_{x,i}}) \notag \\ 
 & \ \ \ \ \le -{{\alpha }_{i3}}(\tau){{\left\| x_i \right\|}^{2}}+{{b}_{i}}({{\alpha }^2_{i4}}(\tau)/\alpha_{i3}(\tau)){{\left\| {{\Delta }_{x,i}} \right\|}^{2}},\notag  \\ 
 & \left\| {\partial {{V}_{i}}}/{\partial x_i}\; \right\|\le {{\alpha }_{i4}}(\tau)\left\| x_i \right\|,\notag \\ 
 & {{\alpha }_{i3}}(\tau)={{\alpha }_{i4}}(\tau)\times {{\alpha }_{i5}}(\tau)   
\end{align}
where ${\alpha}_{ij}(\cdot)$, $j=1,\ldots,5$ are {class $\mathcal{K}_\infty$ functions}, $b_i>0$.

(ii) Assumption on interconnection $g_i(t,x_{i},x_{i+1})$, $i=1,\cdots, n-1$:   there exist a class $\mathcal{K}_\infty $ function  $\alpha_{i6}(\cdot)$ and  a positive constant $c_{g,i}$ satisfying
\begin{equation} \label{assumption_g}
\begin{aligned}
&\left\| ({\partial {{V}_{i}}}/{\partial x_i}\; ) g_i(t,x_{i},x_{i+1}) \right\| \le {{\alpha }_{i6}}(\tau)c_{g,i}\left\| x_i \right\|, \\
& {{\lim }_{t\to {{t}_{f}}}} {{{\alpha }_{i6}} (\tau)/ (\alpha_{i3}(\tau) \alpha_{i+1,5}(\tau))}=0.
\end{aligned}
\end{equation} 
\end{theorem}

\begin{Proof}
See Appendix B.
\end{Proof}

Here we give an example which has been introduced in \cite{song2017time} for justification of assumption (\ref{assumption_cascaded}).
Consider the first-order system governed by
$\dot{x}=b(x,t)u+f(x,t)$
where $x,u \in \mathcal{R}$ and $0<\underline{b}\le \left| b(x,t) \right|$, $\left| f(x,t) \right|\le d(t)\psi (x)$. Employing the residual-time-based scaling function $\mu (t-{{t}_{0}})=\frac{{{T}^{1+m}}}{{{(T+{{t}_{0}}-t)}^{1+m}}},t\in [{{t}_{0}},{{t}_{0}}+T]$ with positive integer $m$, the controller was derived 
$u=-\frac{1}{{\underline{b}}}(k+\lambda \psi {{(x)}^{2}}+\frac{1+m}{T})\mu (t-{{t}_{0}})x$ with $k, \lambda>0$. 
Following the Lyapunov function candidate $V={{{\omega }^{2}}}/{2}\;$, $\omega =\mu (t-{{t}_{0}})x$ and the presentation in \cite{song2017time}, upon applying the control law, one obtains
$\dot{V}\le -2k\mu V+\frac{\mu }{4\lambda }{{d}^{2}}$.
In accordance with the proposed assumption, it is obvious
${{\alpha }_{i1}}(\tau )={{a}_{1}}{{\mu }^{2}}$$(0<{{a}_{1}}<0.5)$, ${{\alpha }_{i2}}(\tau )={{a}_{2}}{{\mu }^{2}}$$({{a}_{2}}>0.5)$,
${{\alpha }_{i3}}(\tau )=k{{\mu }^{3}}$, ${{\alpha }_{i4}}(\tau )=\mu^2 $, ${{\alpha }_{i5}}(\tau )=k{{\mu }}$, ${{\alpha }_{i6}}(\tau )=\mu$, ${{b}_{i}}={1}/{(4\lambda }\;)$,
all of which satisfy the condition in the applied assumption. 

\begin{remark} \label{Remark_strict}
A strict feedback form of the cascaded system (\ref{EQ_Cascaded_N})  can be obtained through the backstepping procedure 
\begin{equation} 
\begin{aligned}
 & {{\Sigma }_{i}}:\dot{x}_i(t)=-k_i(\tau)x_i+g_i(t,x_i,x_{i+1})x_{i+1}+{{\Delta }_{x,i}}, \\
  & {{\Sigma }_{n}}:\dot{x}_n(t)=-k_n(\tau)x_n+{{\Delta }_{x,n}},  i=1,\cdots,n-1
\end{aligned}
\end{equation} 
where ${k}_{i}(\cdot)$, $i=1,\ldots,n$ are class $\mathcal{K}_\infty$ functions. 
{ In principle, each subsystem can achieve the prescribed-time convergence due to the high-gain nature of $k_i(\tau)$ at the prescribed time $t_f$. Under the premise of (\ref{assumption_g}), since the convergence of $\Sigma_i$ is influenced by $\alpha_{i4} \Vert x_{i}\Vert \Vert x_{i+1} \Vert$ or $\alpha_{i4} \Vert x_{i}\Vert ^2 \Vert x_{i+1} \Vert$, an ideal manner would be for $x_{i+1}$ to converge to zero before $\alpha_{i4} $ approaches infinity, which inspires the design $k_{i+1}(\tau)>k_{i}(\tau)$.}
\end{remark}

\begin{remark}
In the high-order prescribed-time control proposed
in \cite{song2017time}, the system model is in a standard integral form
which limits that no uncertainties occur in the  $x_i$'s dynamics. The reason lies in the repeated differentiation of the time-scaling transformation $\mu(t)x_1$ where $\mu(t)$ is a time-varying function which goes to infinity when $t$ approaches $t_f$, as the form.  
 However, the proposed cascaded prescribed-time control scheme in Theorem \ref{Theorem_Cascade_N} has the distinguished features:
\begin{enumerate} [(i)]
\item  Eliminate restrictions on the system model, and allow the uncertainties in $x_i$' dynamics; 
\item {Allow the remaining-time ($t_f-t$)-related interconnection $g_i(t,x_{i},x_{i+1})$ appear in $x_i$' dynamics; }
\item Each subsystem can also be individually designed for its prescribed-time stability performance.
\end{enumerate} 
\end{remark}
\subsection{Prescribed-time Controller design}
For the sake of clarity, the following prescribed-time control design borrows from the techniques of backstepping and the overall dynamics are organized into a cascaded form.
Invoking the definition of $\tau(t,t_f)$ in Definition \ref{definition_2}, let a user-defined $\tau(t,t_f)$ satisfy $\tau(0,t_f)=1$.
Introduce now a time-varying scaling function of $t_f-t$ with 
${{h}}>0$ 
{
\begin{equation}\label{Eq_3_9}
{{\mu }}(t,h)=\left\{ \begin{matrix}
  \tau^h(t,t_f), \ \ \ t\in \left[ 0,{{t}_{f}} \right)   \\
   1, \ \ \ \ \ \ \ \ \ \ \ \ t\in \left[ {{t}_{f}},\infty  \right)    \\
\end{matrix} \right. 
\end{equation}
whose derivative with respect to $t$ yields
\begin{equation} 
\begin{aligned}
& {\dot{\mu}}(t,h) =\left\{ \begin{matrix}
 h{{\mu }}(t,h)(\dot{\tau}/\tau), \ \ t\in \left[ 0,{{t}_{f}} \right)   \\
   0,   \ \ \ \ \ \ \ \ \ \ \ \ \ \ \ \ \ \ t\in \left[ {{t}_{f}},\infty  \right)    \\
\end{matrix} \right. \\
& \dot{\tau}(t,t_f)=0, \ \ \ \ \ \ \ \ \ \ \ \ \ \ \ \ \ \ \  t\in \left[ {{t}_{f}},\infty  \right)
\end{aligned} 
\end{equation}
For simplicity, define $\mu_1(t)=\mu(t,h_1)$, $\mu_2(t)=\mu(t,h_2)$ with $h_1, h_2>0$.}

{
\begin{theorem} \label{Theorem_Kinematics}
Consider the kinematic model (\ref{Eq_3_2}) (\ref{Eq_3_3}). Assume there exists a positive $\sigma $ satisfying ${{ {F}}^{T}} {F} \le  \sigma^2$, $\sigma>0$ due to the boundedness of $d_V$ and $d_{\theta}$. Choose the controller 
\begin{subequations}
\begin{align}
& \left[ \begin{matrix}
   \alpha_{V}^{*}  \\
   \alpha_{\eta}^{*}  \\
\end{matrix} \right]=  -({{k}_{1}} \mu_1 + \frac{h_1 \dot{\tau}} {\tau}) \left[ \begin{matrix}
   {{R}_{e}}  \\
   R{{q}_{e}}  \\
\end{matrix} \right]-{G}  +{M} \left[ \begin{matrix}
   {{V}_{d}}  \\
   {{\eta}_{d} }  \\
\end{matrix} \right],  \\
& \left[ \begin{matrix}
   {V}_{s}^{*}  \\
   {\eta }_{s}^{*}  \\
\end{matrix} \right]=\left\{ \begin{matrix}
{{M}^{-1}} \left[ \begin{matrix}
   \alpha_{V}^{*}  &
   \alpha_{\eta}^{*}  \\
\end{matrix} \right]^T, \ \ \ t  \in \left[ 0,t_f  \right)   \\
f_{M} (\left[ \begin{matrix}
   \alpha_{V}^{*}  &
   \alpha_{\eta}^{*}  \\
\end{matrix} \right]), \ \ \ \ \ \ t \in \left[ t_f, \infty \right)    \\
\end{matrix} \right. \label{Eq_3_10} 
\end{align}
\begin{align}
& \left[ \begin{matrix}
   {{u}_{V}}  \\
   {{u}_{\theta }}  \\
\end{matrix} \right]={{{B}}^{-1}} \left\{ -(({{k}_{2}}+{{\lambda }_{2}}){{\mu }_{2}}+\frac{h_2\dot{\tau}}{\tau})\left[ \begin{matrix}
   {{e}_{V}}  \\
   {{e}_{\eta }}  \\
\end{matrix} \right]-\left[ \begin{matrix}
   \dot{V}_{s}^{*}  \\
   \dot{{\eta }}_{s}^{*}  \\
\end{matrix} \right]-H\right\} \label{Eq_3_16}
\end{align}
\end{subequations}
where ${{k}_{1}}>0$, ${{k}_{2}}>0,{{\lambda }_{2}}>0$ and the function
\[ f_{M} (\left[ \begin{matrix}
   \alpha_{V}^{*}  &
   \alpha_{\eta}^{*}  \\
\end{matrix} \right]) =\left[ \begin{matrix}
\Vert \left[ \begin{matrix}
   \alpha_{V}^{*}  &
   \alpha_{\eta}^{*}  \\
\end{matrix} \right]\Vert    \\
 \text{atan}(-\alpha_{V}^{*}/\alpha_{\eta}^{*})    \\
\end{matrix} \right]. \]
 Define the tracking errors  ${{e}_{V}}={{{V}}_{s}}- {V}_{s}^{*}$, ${{e}_{\eta }}={{{\eta }}_{s}}-{\eta }_{s}^{*}$. Consequently, the closed-loop system  is prescribed-time stable, which means, $({{R}_{e}},{{q}_{e}},e_{V},e_{\eta})$ converges to the origin before $t=t_f$. Furthermore, one obtains
\begin{enumerate}[(i)]
\item when $R_f=0$, the choices of $\mu_1(t)$ and $\mu_2(t)$ should satisfy ${{\lim }_{t\to {{t}_{f}}}}  (t_f-t){ \mu_1 \mu_2}=\infty$;
\item when $R_f>0$,  the formation gets maintained in $(R_f,q_f)$ through $V_{df}=0$, $\dot{\eta}_{d}=0$ and $\dot{V}_{d}=0$ for $t \ge t_f$.
\end{enumerate}  
\end{theorem}}

\begin{Proof}
The proof is divided into four step:

\emph{Step 1.} Prescribed-time convergences of $R_e$ and $q_e$ under $V_s^*$ and $\eta_s^*$ in (\ref{Eq_3_10}) before $t=t_f$.

{
With the controller (\ref{Eq_3_10}), the error dynamics is obtained as
\begin{equation}\label{Eq_3_11}
 \left[ \begin{matrix}
   {\dot{R}_{e}}  \\
   {\dot{q}_{e}}  \\
\end{matrix} \right]=-({{k}_{1}} \mu_1 + \frac{h_1 \dot{\tau}} {\tau}) ) \left[ \begin{matrix}
   {{R}_{e}}  \\
   {{q}_{e}}  \\
\end{matrix} \right].
\end{equation}
Consider the Lyapunov candidate ${{V}_{1}}(r)=x_{1}^{T}{{x}_{1}}$ and ${{W}_{1}}(r)=\mu _{1}^{2}x_{1}^{T}{{x}_{1}}=\omega _{1}^{T}{{\omega }_{1}}$ with ${{x}_{1}}={{\left[ \begin{matrix}
   {{R}_{e}} & {{q}_{e}}  \\
\end{matrix} \right]}^{T}}$ and ${{\omega }_{1}}={{\mu }_{1}}{{\left[ \begin{matrix}
   {{R}_{e}} & {{q}_{e}}  \\
\end{matrix} \right]}^{T}}$. The derivative of ${{W}_{1}}(r)$ along (\ref{Eq_3_11}) is
\begin{equation}\label{Eq_3_12}
\begin{aligned}
 {\dot{W}_{1}}(t) & =2\omega _{1}^{T}\left[   {\dot{\mu }_{1}}  {{x}_{1}}-({{k}_{1}} \mu_1 + {\dot{\tau}}/{\tau}){{\mu }_{1}}{{\omega }_{1}} \right] \\
 & =2\omega _{1}^{T}\left[  h_1(\dot{\tau}/\tau)  {{\omega}_{1}}-({{k}_{1}} \mu_1 + h_1 {\dot{\tau}}/{\tau}) {{\omega }_{1}} \right] \\
 &\le -2{{k}_{1}}{{\mu }_{1}}{{W}_{1}}.
 \end{aligned}
\end{equation}}
Solving this differential inequality  and invoking the fact $\int_0^t{\mu_1(\nu)\text{d}\nu}=-\phi_1(t)$ give
$ \mu _{1}^{2}(t){{V}_{1}}(t)<{{\exp }^{2{{k}_{1}}{{\phi }_{1}}}}{{W}_{1}}(0)$
which yields 
\begin{align} \label{Eq_3_10a} 
& \left\| \left[ \begin{matrix}
   {{R}_{e}}  \\ {{q}_{e}}  \\
\end{matrix} \right] \right\|\le \frac{1}{{{\mu }_{1}}(t)}{{\exp }^{{{k}_{1}}{{\phi }_{1}}}}(R_{e}^{2}(0)+q_{e}^{2}(0)).  
\end{align}
where ${{\phi }_{1}}(t)=\frac{{{{t}}_{f}}}{{{h}_{1}}-1}(1-\mu _{1}^{{({{h}_{1}}-1)}/{{{h}_{1}}}\;})$. Thus, the origin of system (\ref{Eq_3_11}) is globally prescribed-time stable within the prescribed time ${{t}_{f}}$. 

\emph{Step 2.} Prescribed-time convergences of $ {V}_s$ to $ {V}_{s}^{*}$ and $ {\eta}_s$ to ${\eta}_{s}^{*}$ under the proposed $u_V$ and $u_\theta$ in (\ref{Eq_3_16}) before $t=t_f$.

{
With the controller (\ref{Eq_3_16}), the dynamic model of ${{e}_{V}}$, ${{e}_{\eta }}$ is obtained as
\begin{equation}\label{Eq_3_18}
 \left[ \begin{matrix}
   {\dot{e}_{V}}  \\
   {\dot{e}_{\eta }}  \\
\end{matrix} \right]=-(({{k}_{2}}+{{\lambda }_{2}}){{\mu }_{2}}+ \frac{h_2 \dot{\tau}}{\tau}) \left[ \begin{matrix}
   {{e}_{V}}  \\
   {{e}_{\eta }}  \\
\end{matrix} \right]+ {F}.
\end{equation}
Consider the Lyapunov candidate ${{V}_{2}}(t)=x_{2}^{T}{{x}_{2}}$ and ${{W}_{2}}(t)=\mu _{2}^{2}x_{2}^{T}{{x}_{2}}=\omega _{2}^{T}{{\omega }_{2}}$ with ${{x}_{2}}={{\left[ \begin{matrix}
   {{e}_{V}} & {{e}_{\eta }}  \\
\end{matrix} \right]}^{T}}$ and ${{\omega }_{2}}={{\mu }_{2}}{{\left[ \begin{matrix}
   {{e}_{V}} & {{e}_{\eta }}  \\
\end{matrix} \right]}^{T}}$.}
By applying Young's inequality with ${{\lambda }_{2}}>0$, one obtains
$\omega _{2}^{T}{{\mu }_{2}}  {F}\le {{\lambda }_{2}}{{\mu }_{2}}\omega _{2}^{T}{{\omega }_{2}}+ {{{\mu }_{2}}{{{  {F}}}^{T}} {F}}/({4{{\lambda }_{2}}})$.
In this way, the derivative of ${{W}_{2}}(t)$ with respect to the variable $t$ yields
\begin{align} \label{Eq_3_20}
{\dot{W}_{2}}(t)&=2\omega _{2}^{T}\left[   
\dot{\mu}_2{{x}_{2}}-(({{k}_{2}}+{{\lambda }_{2}}){{\mu }_{2}}+h_2\dot{\tau}/\tau){{\omega }_{2}}+{{\mu }_{2}}  {F} \right] \notag \\ 
 & \le -2{{k}_{2}}{{\mu }_{2}}{{W}_{2}}+ {{{\mu }_{2}}{{{  {F}}}^{T}} {F}}/({2{{\lambda }_{2}}}).  
\end{align}

In view of ${{ {F}}^{T}} {F} \le  \sigma^2$ and invoking Lemma 1 in \cite{song2017time}, it is direct to deduce 
$\mu_2^2(t){{V}_{2}}(t)< \left({{\exp }^{2{{k}_{2}}{{\phi }_{2}}}}{{W}_{2}}(0)+ {\sigma^2}/({4k_2 \lambda_2})\right)
$
which yields
\begin{align}  
\left\| \left[ \begin{matrix}
   {{e}_{V}} \\  {{e}_{\eta}}  \\
\end{matrix} \right] \right\|\le \frac{1}{{{\mu }_{2}}}\left( {{\exp }^{{{k}_{2}}{{\phi }_{2}}}}(e_{V}^{2}(0)+e_{\theta}^{2}(0))+ \frac{\sigma}{2\sqrt{k_2 \lambda_2 }}\right) \label{Eq_3_17}
\end{align}
where ${{\phi }_{2}}(t)=\frac{{{{t}}_{f}}}{{{h}_{2}}-1}(1-\mu _{2}^{{({{h}_{2}}-1)}/{{{h}_{2}}}\;})$. Thus, it follows the prescribed-time stability of $(e_V,e_{\eta})$.

\emph{Step 3:} Prescribed-time convergence of the cascaded system.

{Combining the two system models into a cascaded form, we obtain 
\begin{subequations}\label{Eq_3_26}
\begin{align}
  &  \left[ \begin{matrix}
   {\dot{R}_{e}}  \\
   {\dot{q}_{e}}  \\
\end{matrix} \right]=-({{k}_{1}} \mu_1 + \frac{h_1 \dot{\tau}} {\tau}) ) \left[ \begin{matrix}
   {{R}_{e}}  \\
   {{q}_{e}}  \\
\end{matrix} \right]+\bar{M}\left[ \begin{matrix}
   {{e}_{V}}  \\
   {{e}_{\eta }}  \\
\end{matrix} \right], \\ 
 &  \left[ \begin{matrix}
   {\dot{e}_{V}}  \\
   {\dot{e}_{\eta }}  \\
\end{matrix} \right]=-(({{k}_{2}}+{{\lambda }_{2}}){{\mu }_{2}}+ \frac{h_2 \dot{\tau}}{\tau}) \left[ \begin{matrix}
   {{e}_{V}}  \\
   {{e}_{\eta }}  \\
\end{matrix} \right]+ {F}.  
\end{align}
\end{subequations}
\begin{enumerate}[(i)]
\item Under the case $R_f>0$,  it is obvious that $\Vert \bar{M} \Vert$ is bounded. In addition, $\alpha_{13}(\tau)=2k_1\mu_1^3$, $\alpha_{14}(\tau)=2\mu_1^2$,  $\alpha_{15}(\tau)=k_1 \mu_1$; $\alpha_{23}(\tau)=2k_2 \mu_2^3$, $\alpha_{24}(\tau)=2\mu^2_2$, $\alpha_{25}(\tau)=k_2 \mu_2$.
Invoking the proposed Theorem \ref{Theorem_Cascade_N}, the cascaded system is globally prescribed-time stable in the sense that the reference trajectory gets tracked before $t_f$;
\item Under the case $R_f=0$,  it is obvious that $\Vert \bar{M} \Vert$ is  related to $1/(t_f-t)$ such that $\alpha_{16}(\tau)=\mu^2_1 /(t_f-t)$. 
Invoking the proposed Theorem \ref{Theorem_Cascade_N}, the cascaded system is globally prescribed-time stable if only  ${{\lim }_{t\to {{t}_{f}}}} \frac{ 1/(t_f-t)}{ \mu_1 \mu_2}=0$.
\end{enumerate}} 

{\emph{Step 4:} Formation keeping over $[t_f, \infty)$ when $R_f>0$.}

{
When $t \in [t_f, \infty)$, $\mu_1=\mu_2=1$, $\dot{\tau}=0$ such that the cascaded system is rewritten into
\begin{subequations} 
\begin{align}
  &  \left[ \begin{matrix}
   {\dot{R}_{e}}  \\
   {\dot{q}_{e}}  \\
\end{matrix} \right]=-{{k}_{1}} \left[ \begin{matrix}
   {{R}_{e}}  \\
   {{q}_{e}}  \\
\end{matrix} \right]+\bar{M}_m \left[ \begin{matrix}
   {{e}_{V}}  \\
   {{e}_{\eta }}  \\
\end{matrix} \right], \\ 
 &  \left[ \begin{matrix}
   {\dot{e}_{V}}  \\
   {\dot{e}_{\eta }}  \\
\end{matrix} \right]=-({{k}_{2}}+{{\lambda }_{2}}) \left[ \begin{matrix}
   {{e}_{V}}  \\
   {{e}_{\eta }}  \\
\end{matrix} \right]+ {F}   
\end{align}
\end{subequations}
where $V_d(t)=0$, $\eta_d(t)=0$, $t \ge t_f$ and
$ \bar{M}_m=\text{diag}(1,1/R)  {M}(R,{{\eta }_{s}},e_\eta,V_s)$.   
Define the Lyapunov candidate $V=x_1^T x_1+ x_2^T x_2$ whose time derivative yields
\begin{equation}
\dot{V} \le -(2k_1 -\Vert \bar{M}_m \Vert)x_1^T x_1 -(2k_2 -\Vert \bar{M}_m \Vert) x_2^T x_2
\end{equation}
where $k_1, k_2 >0.5 \Vert \bar{M}_m \Vert$.
The corresponding stabilization is obvious established according to the above proof.}
\end{Proof}

\begin{remark}
It is obvious that the larger $k_i$, $h_i$ ($i=1,2$) are, the faster $R_e$, $q_e$ converge. Invoking Remark \ref{Remark_strict},  it is better to design $k_2 \mu_2 >k_1 \mu_1$  to make the states of different subsystems converge in a reasonable order. { In addition,
the high-order scaling of state by a function of time in \cite{song2017time} requires the parameter $h$ to be positive integers greater than 1. Our results do not require such limitations.}
\end{remark}

\section{Extensions to simultaneous tracking/formation of multi-vehicles}
For the multi-agent tracking/formation scenario, a distinguishing requirement is the simultaneous arrival. Suppose that $n$ vehicles, $M_1, \cdots, M_n$, participates in a tracking/formation with respect to a single target. In the following, the variable $X_i, 1 \le i \le n$ stands for the corresponding $X$ of the $i$-th vehicle. With the above design and analysis, each vehicle realizes the prescribed-time stabilization, that is, it realizes the tracking of $R_i$ to $R_{d,i}$ and $q_i$ to $q_{d,i}$ before the terminal time $t_{f,i}$. 

According to the proposed reference relative tracking trajectory theorem in Section II.A, for the simultaneous tracking/formation with $t_{f,i}=T_d$, the sufficient condition is
$\int_{0}^{T_d}{{{V}_{d,i}}(t)\text{d}t}=S_{d,i}$, $i=1,\cdots,n$.

Assume a centralized coordination strategy exists between vehicles and the desired terminal time can be determined at the initial time. Thus, it is a direct solution to allocate the velocity reasonably to obtain simultaneous arrival. Below is an adjustment strategy which can achieve the smooth arrival to the terminal situation.

Construct a $m$-order polynomial function to describe the law of ${{V}_{d,i}}(t)$
\begin{equation}\label{Eq_4_4}
{{V}_{d,i}}(t)= I_1 t^m+I_2 t^{m-1}+\cdots+I_{m} t+I_{m+1}
\end{equation}
which satisfies $m+1$ equality constraints, including the following 3 necessary conditions
\[{{V}_{d,i}}(0)=V_{d0,i} ,  {{V}_{d,i}}(T_d)=V_{df,i}, \int_0^{T_d}{{{V}_{d,i}}}(\tau)\text{d}\tau =S_{d,i}. \]

%\begin{remark}
%When it takes into consideration the distributed collaborative strategy,  a communication graph exists between vehicles. The coordination strategy balances the terminal time $t_{f,i}$ of each vehicle through the adjustment of their respective velocities.
%\end{remark}

\begin{figure}[t]
\centerline{\includegraphics[width=0.47\textwidth]{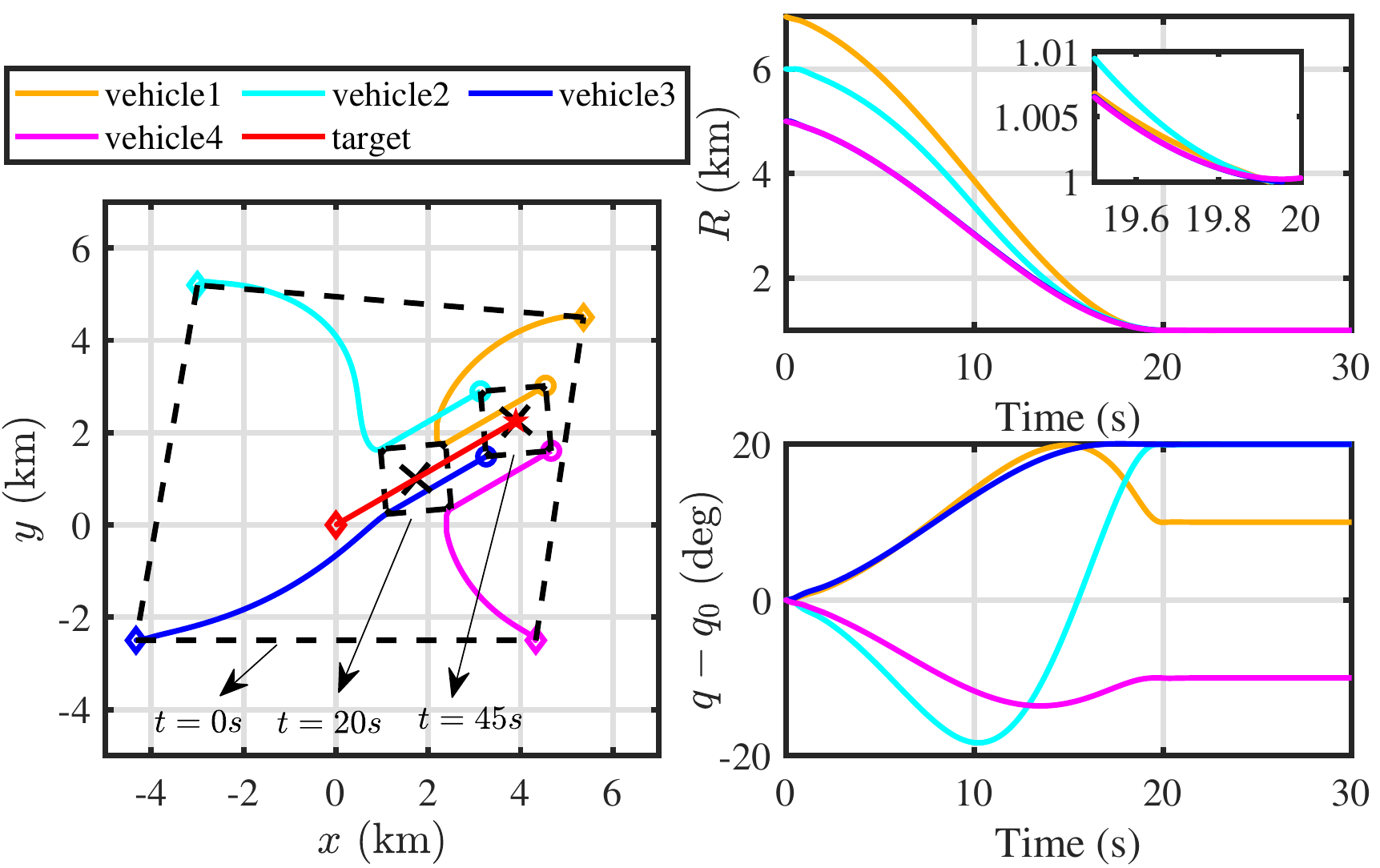}}
\caption{Trajectories of four vehicles}
\label{fig:Fig_whole_movingtarget}
\end{figure}

\begin{figure}[tbp]
\centerline{\includegraphics[width=0.5\textwidth]{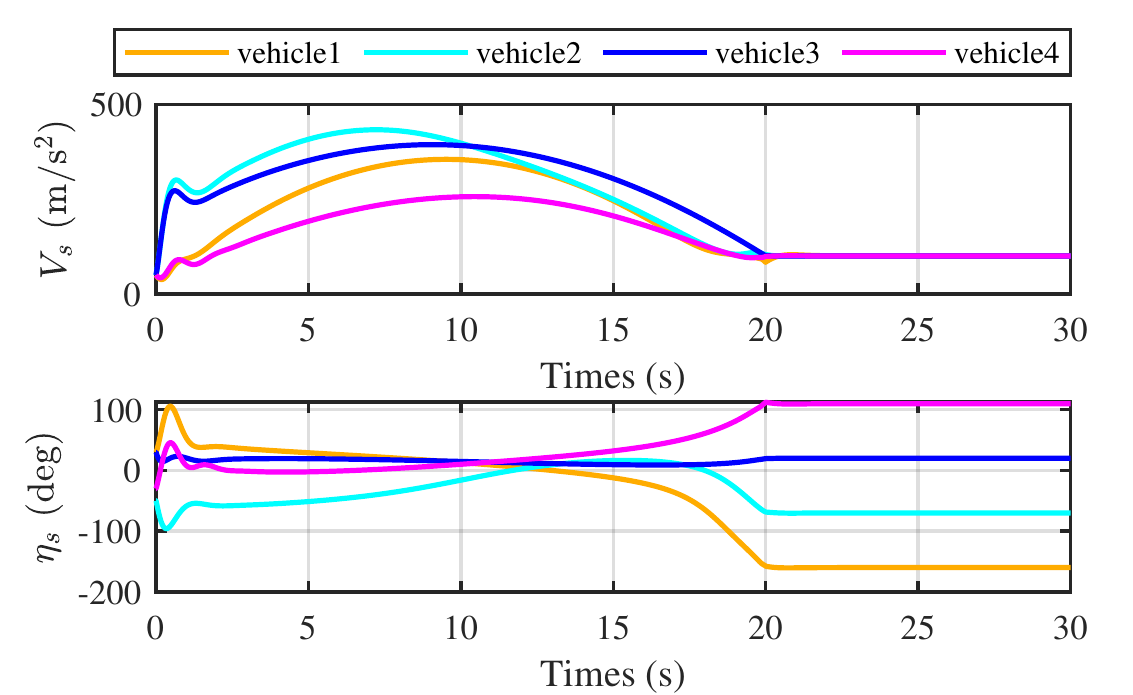}}
\caption{States of four vehicles}
\label{fig:Fig_vehicle_movingtarget}
\end{figure}

\section{Numerical examples}

\begin{table}[bp] \footnotesize 
 \newcommand{\tabincell}[2]{\begin{tabular}{@{}#1@{}}#2\end{tabular}}
 \caption{Scenarios for simultaneous formation.}
 \label{Table_5_1}
 \centering
%\renewcommand\arraystretch{1.5}
%\linespread{5.5}
 \setlength{\tabcolsep}{0.2mm}{
 \begin{tabular}{ccccccc}%
  \toprule
  %\hline
  \multirow{3}{*}{} 
  &  \multicolumn{3}{c@{}@{}@{}}{Initial condition} & \multirow{3}{*}{} 
  & \multicolumn{2}{@{}c@{}}{Terminal condition} \\[2.5pt]
  \cmidrule{2-4} \cmidrule{6-7}
  %\cline{4-5}
  & Distance $R_0$ & LOS $q_0$ &Path angle $\theta_0$ & & Distance $R_f$  & LOS $q_f$  \\
  \midrule
 %\hline
  Vehicle 1 & 7000 m  &  220 deg &  190 deg  &   & 1000 m & 230 deg \\
  Vehicle 2 & 6000 m  &  -60 deg &  -10 deg  &   & 1000 m & -40 deg \\
  Vehicle 3 & 5000 m  &  30 deg  &  0   deg  &   & 1000 m & 50 deg \\
  Vehicle 4 & 5000 m  &  150 deg &  180 deg  &   & 1000 m & 140 deg \\
  \bottomrule
 %\hline
 \end{tabular}}
\end{table} 

Suppose that four vehicles track a single target with the initial and terminal condition shown in Table \ref{Table_5_1}. Each vehicle has the same initial velocity, 50 m/s. 

\begin{figure}[t]
\centerline{\includegraphics[width=0.5\textwidth]{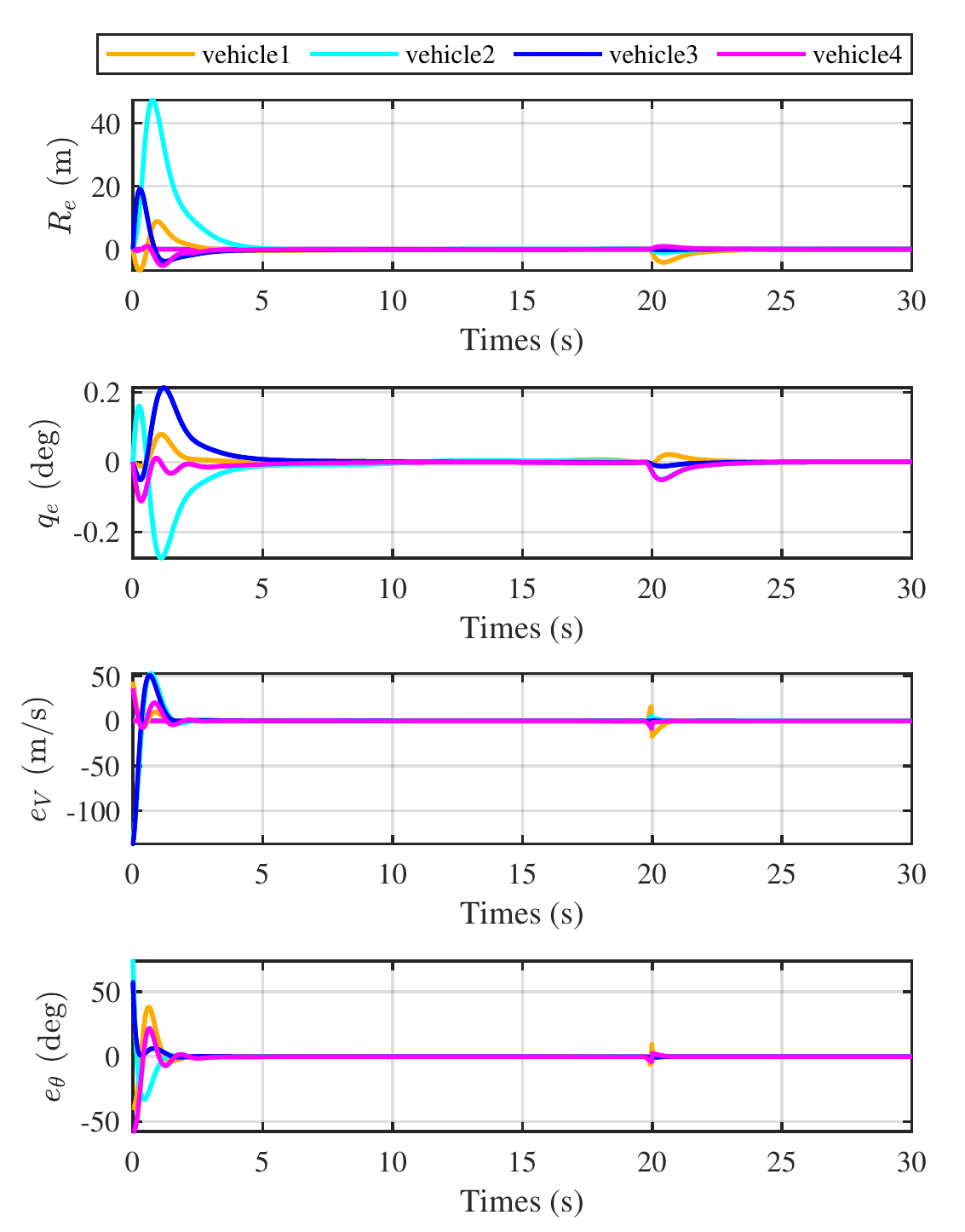}}
\caption{Tracking errors of four vehicles}
\label{fig:Fig_error_movingtarget}
\end{figure}

\begin{figure}[t]
\centerline{\includegraphics[width=0.5\textwidth]{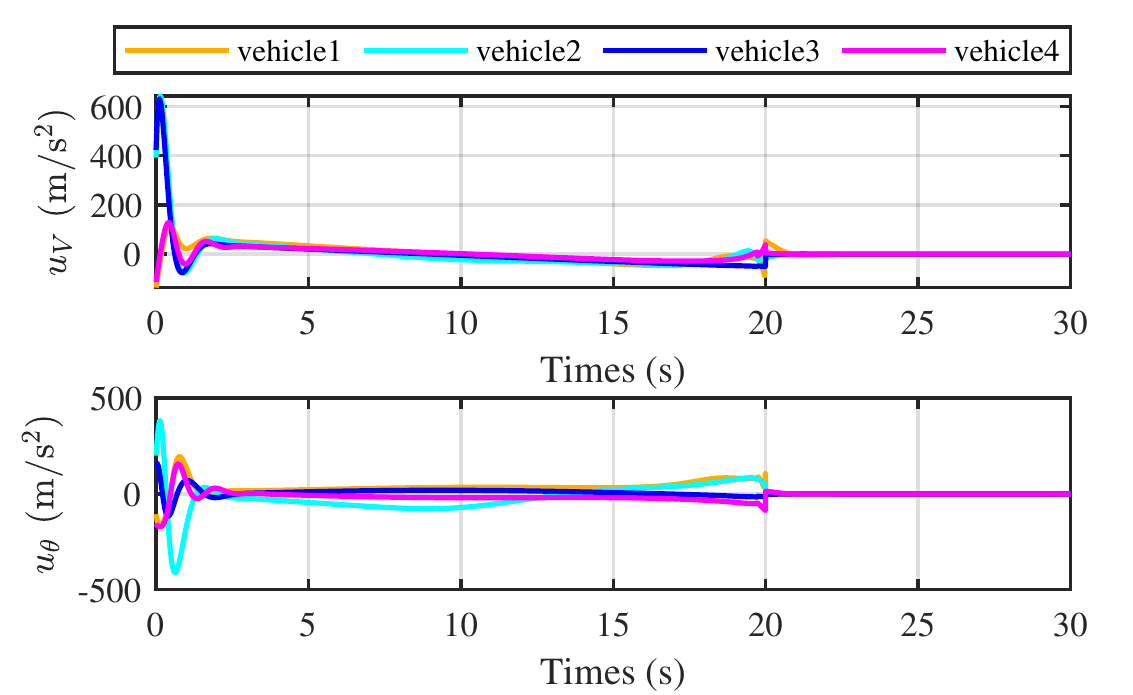}}
\caption{Control input of four vehicles}
\label{fig:Fig_control_movingtarget}
\end{figure}

For a target moving at a speed of $V_t=100$m/s in a 45-degree direction, the scenario at $T_d=t_{f,i}=20s$ expects the four vehicles to be at a 90-degree angle to each other and a distance of 1000m from the target, meanwhile, it requires  $V_{df,i}=0$m/s for the gentle transition to the phase of formation keeping $t>T_d$.

In accordance with (\ref{Eq_4_4}), use a third-order polynomial for velocity planning. { The function $\tau(t,t_f)$ is chosen in form of $\tau(t,t_f)=1+\text{ln}(\frac{t_f}{t_f-t})$.} The control parameters are as follows:
$N_1=6$, $N_2=10$, $k_1=1$, $k_2=2$, $h_1=1$, $h_2=2$ and $\lambda_2=1$.
The corresponding results for Case 2 can be found in Fig. \ref{fig:Fig_whole_movingtarget}-\ref{fig:Fig_control_movingtarget}. {From the curves of $R_i$ and $q_i$ in Fig. \ref{fig:Fig_whole_movingtarget}, 
the simultaneous formation and formation keeping are obvious achieved at $T_d$, and the transmission process to formation keeping is 
smooth due to the small value of the desired relative velocity $V_{df,i}$.}
The prescribed-time convergence of tracking errors can be obtained in Fig. \ref{fig:Fig_error_movingtarget}.
It is observed from Fig. \ref{fig:Fig_vehicle_movingtarget} that the changes of both velocities and angles are slow towards the prescribed time $T_d$,  which promotes the limited amount of control input near the time $T_d$ as shown in Fig. \ref{fig:Fig_control_movingtarget}. 
After the prescribed time $T_d$, the task becomes formation keeping relative to the target.
The controller switching brings the controller mutation at time $T_d$ in Fig. \ref{fig:Fig_control_movingtarget}. As shown in the short period after $T_d$ in Fig. \ref{fig:Fig_error_movingtarget}, 
there is a regulation stage for the angle $\eta_{s,i}$ such that the speed of each vehicle gets to be parallel to that of the target.

It is worth emphasizing that the realization of simultaneous tracking and formation is completely dependent on the proposed space-and-time-synchronized strategy.

\section{Conclusion}
In this paper, we presented a space-and-time-synchronized control method for simultaneous tracking/ formation. The resultant control is able to achieve the predetermined state at the prescribed terminal time with a fixed relative space trajectory independent of time. As a result, simultaneous tracking/formation of multiple vehicles can be directly implemented. Extending this method to the simultaneous tracking/formation with no predetermined terminal time under communication topologies is an interesting topic for future research.

\appendices
\section{Proof of Theorem 1}
\emph{Step 1.} Prescribed-distance stability before $r_d=r_f$.

{Similar to the prescribed-time design philosophy in \cite{PWang2021prescribed}, we choose the controller (\ref{Eq_2_6}) as a prescribed-distance controller by treating $r_f-r_d$ as a measure of the remaining distance to go.} Substitution of (\ref{Eq_2_6}) into (\ref{Eq_2_5}) leads to second-order differential equation  
\begin{equation}\label{Eq_2_7}
\frac{{\text{d}^{2}}{{q}_{d}}}{\text{d}r_{d}^{2}}+\frac{{{k}_{d2}}}{{{{r}}_{f}}-{{{r}}_{d}}}\frac{\text{d}{{q}_{d}}}{\text{d}{{{r}}_{d}}}+\frac{{{k}_{d1}}}{{{({{{r}}_{f}}-{{{r}}_{d}})}^{2}}}({{q}_{d}}-{{q}_{f}})=0.
\end{equation} 
This is a kind of second-order Cauchy equation. Here, we introduce a scaling transformation $\zeta =\ln ({{r}_{f}}-{{r}_{d}})$ 
with which the differential equation is transformed into
\begin{equation}\label{Eq_2_8}
\frac{{\text{d}^{2}}{{q}_{d}}}{\text{d}{{\zeta }^{2}}}-({{k}_{d2}}+1)\frac{\text{d}{{q}_{d}}}{\text{d}\zeta }+{{k}_{d1}}({{q}_{d}}-{{q}_{f}})=0.
\end{equation}
The corresponding characteristic equation is obtained by letting ${{q}_{d}}-{{q}_{f}}={{e}^{\lambda \zeta }}$ as
$
{{\lambda }^{2}}-({{k}_{d2}}+1)\lambda +{{k}_{d1}}=0 $
whose discriminant is 
$\Delta ={{({{k}_{d2}}+1)}^{2}}-4{{k}_{d1}}$. 
In order to obtain an elegant description of the solution, we make ${{\lambda }_{1}}=N_1 \in \mathcal{N}_+$, ${{\lambda }_{2}}=N_2 \in \mathcal{N}_+$, $N_1<N_2$ which brings about ${{k}_{d1}}=N_1N_2, \  {{k}_{d2}}=N_1+N_2-1$, and the discriminant $\Delta >0$. Hence, with the initial condition ${{q}_{d}}(0)={{q}_{0}}$ and $\tan ({{\eta }_{d}}(0))=\tan {{\eta }_{d0}}$, the solution of this Cauchy equation can be obtained
\begin{equation}\label{Eq_2_10}
{{q}_{d}}({{r}_{d}})={{C}_{1}}({{r}_{f}}-{{r}_{d}})^{N_1}+{{C}_{2}}{{({{r}_{f}}-{{r}_{d}})}^{N_2}}+{{q}_{f}}
\end{equation}
where 
\begin{align}
&{{C}_{1}}=\frac{N_2({{q}_{0}}-{{q}_{f}})+{{{r}}_{f}}\tan {{\eta }_{d0}}(R_0-{R}_{f1})/R_0}{(N_2-	N_1){{{r}}^{N_1}_{f}}}, \notag \\
& {{C}_{2}}=\frac{N_1({{q}_{f}}-{{q}_{0}})-{{{r}}_{f}}\tan {{\eta }_{d0}}(R_0-{R}_{f1})/R_0}{(N_2-N_1)r_{f}^{N_2}}. \notag
\end{align}
Differentiating ${{q}_{d}}(r_d)$ with respect to $r_d$ yields
\begin{equation}\label{Eq_2_11}
\begin{aligned}
\tan {{\eta }_{d}}= -\frac{R_d}{R_{d}-{R}_{f1}} &\left [  N_1{{C}_{1}}{{({{r}_{f}}-{{r}_{d}})}^{N_1-1}} \right. \\
&  \ \ \ \left. +N_2{{C}_{2}}{{({{r}_{f}}-{{r}_{d}})}^{N_2-1}} \right].
\end{aligned}
\end{equation}
It is thus clear that, given the positive integers ${{N}_{1}}$, ${{N}_{2}}$,  as $r_d$ approaching $r_f$, ${{q}_{d}}({{r}_{d}}) $ and  ${{\eta }_{d}}({{r}_{d}})$converges to $q_f$ and 0, respectively.

\emph{Step 2.} Position maintaining after $R_d$ approaching $R_f$.

Since $\eta_d(r_f)=0$ in (\ref{Eq_2_11}), $\eta_d(r_d) \equiv 0$  through $\dot{\eta}_d=0$ for $r_d \ge r_f$. When $ {V}_{df}=0$ and $\dot{V}_d=0$ for $r_d \ge r_f$, it is apparent $R_d \equiv R_f$ after $R_d$ approaching $R_f$. Invoking $ {{{\text{d}}^{2}}{{q}_{d}}}/{\text{d}r_{d}^{2}}=0$  in  (\ref{Eq_2_5}), one obtains  $q_d(r_d) \equiv 0$. 
 
\section{Proof of Theorem 2}
We start from the simplified cascaded system with $n=2$ subsystems.
Evidently the prescribed-time stability of $x_2(t)$ in system ${{\Sigma }_{2}}$ can be obtained by taking time derivative of ${{V}_{2}}(t,x_2)$
\begin{equation}\label{Eq_3_6}
\begin{aligned}
   {{{\dot{V}}}_{2}}
 & \le -{{\alpha }_{23}}(\tau){{\left\| x_2 \right\|}^{2}}+{{b}_{2}}({{\alpha }^2_{24}}(\tau)/\alpha_{23}(\tau)){{\left\| {{\Delta }_{x,2}} \right\|}^{2}} \\ 
 & =-{{\delta }_{2}}{{\alpha }_{23}}{{\left\| x_2 \right\|}^{2}} \\
 & \ \ \ -{{\alpha }_{23}}\left[ (1-{{\delta }_{2}}){{\left\| x_2 \right\|}^{2}}-{{b}_{2}}{{{\left\| {{\Delta }_{x,2}} \right\|}^{2}}}/{{{\alpha }^2_{25}}}\; \right]  
\end{aligned}
\end{equation}
with ${{\delta }_{2}}\in (0,1)$. Define the region ${{\Omega }_{x_2}}(r)=\left\{ x_2|{{\alpha }^2_{25}}{{\left\| x_2 \right\|}^{2}}\le {{b}_{2}}{{\left\| {{\Delta }_{x,2}} \right\|}^{2}}/(1-{{\delta }_{2}}) \right\}$ and we get ${{\Omega }_{{x_2}}}({{t}_{f}})={{\lim }_{t\to {{t}_{f}}}}{{\Omega }_{{x_2}}}(t)=0$, following from ${{\lim }_{t\to {{t}_{f}}}}{{\alpha }_{25}}(\tau)=\infty $.  When $x_2$ is outside of ${{\Omega }_{x_2}}(r)$, the increasing property of ${{\alpha }_{23}}(\tau)$ will force $x_2$ to ${{\Omega }_{x_2}}(t)$;  once $x_2$ is inside of ${{\Omega }_{x_2}}(t)$,  $x_2$ will never escape  ${{\Omega }_{x_2}}(t)$ and converge to 0 as  $t$ approaching ${{t}_{f}}$.

   Then, the time derivative of ${{V}_{1}}(t,x_1)$ along ${{\Sigma }_{1}}$ becomes
\begin{align}
  {{{\dot{V}}}_{\text{1}}}  
 & \le -{{\alpha }_{13}}{{\left\| x_1 \right\|}^{2}}+{{b}_{1}}({{\alpha }^2_{14}}/\alpha_{13}){{\left\| {{\Delta }_{x,1}} \right\|}^{2}} +{{\alpha }_{16}} {{c}_{g,1 }} \left\| x_1 \right\|  \left\| x_2 \right\| \notag \\ 
 & =-0.5{{\delta }_{1}}{{\alpha }_{13}}{{\left\| x_1 \right\|}^{2}}-{{\alpha }_{13}}\left[ (1-{{\delta }_{1}}){{\left\| x_1 \right\|}^{2}} \right. \notag \\
 & \ \ \ \ \left. -\gamma \alpha^2_{16} {{{\left\| x_2 \right\|}^{2}}}/ {\alpha^2_{13}}  -{{b}_{1}}{{{\left\| {{\Delta }_{x,1}} \right\|}^{2}}}/{{{\alpha }^2_{15}}}\; \right]  
\end{align}
where $\gamma ={0.5c_{\theta }^{2}}/{{{\delta }_{1}}}\;$. In view of the relationship among $\alpha_{13}$, $\alpha_{16}$ and $\alpha_{25}$, one obtains
\begin{align}
 {{\lim }_{t\to {{t}_{f}}}}{\alpha^2_{16} {{{\left\| x_2 \right\|}^{2}}}/{\alpha^2_{13}}}&= \alpha^2_{16} /(\alpha^2_{13}\alpha^2_{25})  \left\| {{\Delta }_{x,2}} \right\|^2    {{{b}_{2}}/(1-{{\delta }_{2}})}   \notag \\
 &=0. 
 \end{align}
Invoking the boundedness of ${{\left\| {{\Delta }_{x,1}} \right\|}^{2}}$, the prescribed-time convergence of $x_1$ in system ${{\Sigma }_{1}}$ can be proved.

The prescribed-time convergence of cascaded system with $n=2$ is thus concluded. 
From the prescribed-time stability of subsystems $\Sigma_n$ and $\hat{\Sigma}_{n-1}$, the prescribed-time stability of ${\Sigma}_{n-1}$ system is obtained; from the prescribed-time stability of subsystems ${\Sigma}_{n-1}$ and $\hat{\Sigma}_{n-2}$, the prescribed-time stability of system ${\Sigma}_{n-2}$ is obtained; and so on, the prescribed-time stability of ${\Sigma}_{1}$ system can be obtained.
The proof is thus completed. 

\textbf{ }

%\textbf{References}
\small
\bibliographystyle{elsarticle-num-names}  
\bibliography{ref}

\end{document}